\input{aipcheck}

\documentclass[
    ,final            
  ]
  {aipproc}

\layoutstyle{6x9}

\begin{document}

\title{The Spin Distribution of Millisecond X-ray Pulsars}


\author{Deepto Chakrabarty}{
  address={Kavli Institute for Astrophysics and Space Research, \\
           Massachusetts Institute of Technology, Cambridge, MA 02139, USA}
}

\begin{abstract}
 The spin frequency distribution of accreting millisecond X-ray
 pulsars cuts off sharply above 730~Hz, well below the breakup spin
 rate for most neutron star equations of state.  I review several
 different ideas for explaining this cutoff.  There is currently
 considerable interest in the idea that gravitational radiation from
 rapidly rotating pulsars might act to limit spin up by accretion,
 possibly allowing eventual direct detection with gravitational wave
 interferometers.  I describe how long-term X-ray timing of fast
 accreting millisecond pulsars like the 599~Hz source IGR J00291+5934
 can test the  gravitational wave model for the spin frequency limit. 
\end{abstract}

\maketitle


\section{INTRODUCTION}

This proceedings volume (and the meeting it arises from) celebrates
the decadal anniversary of the discovery of the first
accretion-powered millisecond pulsar SAX
J1808.4$-$3658 \citep{wv98,cm98}, and more generally the discovery of
millisecond X-ray variability tracing (or at least coupled to) the
spins of accreting neutron stars in low-mass X-ray
binaries \citep{szs+96,vsz+96,cmm+03}. These discoveries, all made
with the Rossi X-ray Timing Explorer (RXTE; \citep{brs93}), finally
verified the proposal that the old, weak-field neutron stars that
comprise the population of millisecond radio pulsars were spun up
(``recycled'') by sustained accretion \citep{acr+82,rs82}. The
detailed history of these ideas is reviewed by Alpar in these
proceedings \citep{alp09}.

The accreting millisecond X-ray pulsars (AMXPs) are obviously an ideal
laboratory for studying the physics underlying the spin and magnetic
evolution of neutron stars at the end of their lives, and particularly
for understanding the pulsar recycling process in detail.  One
approach is to measure the short-term spin evolution of individual AMXPs
in order to study the action of magnetic accretion torques. Spin
frequency derivatives have been reported during transient outbursts of
several accretion-powered AMXPs, with spin-up/spin-down rates of order 
$\sim 10^{-14}$~Hz~s$^{-1}$ \citep{gcm+02,bdm+06,bdl+07,pmb+08,rdb+08},
consistent with the scale predicted by standard magnetic accretion
torque theory for the observed luminosities \citep{gl79,pc99}.  However,
some pulsars are subject to substantial pulse shape variability over
an outburst, which can potentially mimic spin evolution
(see \cite{hpc+08}).  This conference witnessed significant debate
about how serious this problem is and how to best mitigate it. In this
paper, I discuss a second approach: using the underlying spin
distribution of the entire ensemble of known AMXPs to explore what
governs their spin evolution.

\section{HOW TO INFER THE SPIN DISTRIBUTION}

How fast can recycled pulsars spin?  We certainly expect a strict
upper limit from centrifugal break-up, with the spin frequency where
this occurs depending upon the equation of state for the ultradense
matter in the neutron star core.  A firm upper limit of $\sim$3~kHz is
set by the requirement that the sound speed in the neutron star
interior not exceed the 
velocity of light, while the most ``favored'' models for the equation
of state have maximum spins in the 1500--2000~Hz
range \citep{cst94,hlz99,lp01}.  

{\bf Radio pulsars; accretion-powered AMXPs.} Ideally, one would use
the large population ($\sim200$) of known millisecond radio pulsars to
infer the underlying spin distribution of recycled pulsars.  However,
there are a number of significant selection biases against the
detection of the fastest millisecond radio pulsars (see,
e.g., \citep{hrs+07}), making it difficult to accurately estimate the
underlying population from the observed sample.  What about the X-ray
systems?  Among the AMXPs, we must distinguish between the {\em
accretion-powered} pulsars (in which the persistent accretion flux is
modulated due to magnetic channeling) and the {\em nuclear-powered}
pulsars (in which oscillations are detected during thermonuclear X-ray
bursts originating at a discrete point on the stellar surface).  Seven
of the ten known accretion-powered millisecond X-ray pulsars are
low-luminosity X-ray transients with short orbital periods (and thus
significant Doppler smearing of the pulsation), raising multiple
concerns about selection biases against detection, as well as the
possibility that the observed population represents only a narrow (and
perhaps unrepresentative) evolutionary subset of the low-mass X-ray
binary population.  The other three accretion-powered millisecond
pulsars show only intermittent pulsations with low duty
cycle \citep{gmk+07,gss+07,acp+08,cap+08}, again raising questions
about a selection bias against detection.

{\bf Nuclear-powered AMXPs.}
On the other hand, the nuclear-powered millisecond X-ray pulsars (the
burst oscillation sources) are an ideal probe of the spin
distribution.  Their oscillations (and their bursts) are bright and
easily detected throughout the Galaxy, the signals are sufficiently
short-lived to make orbital Doppler smearing irrelevant, their host
systems sample a wide variety of orbital periods and evolutionary
histories among the low-mass X-ray binaries, and the RXTE/PCA
instrument has no sigificant selection biases against detecting
oscillations as rapid as $\sim$2~kHz.  

\begin{figure}
  \includegraphics[height=.45\textheight]{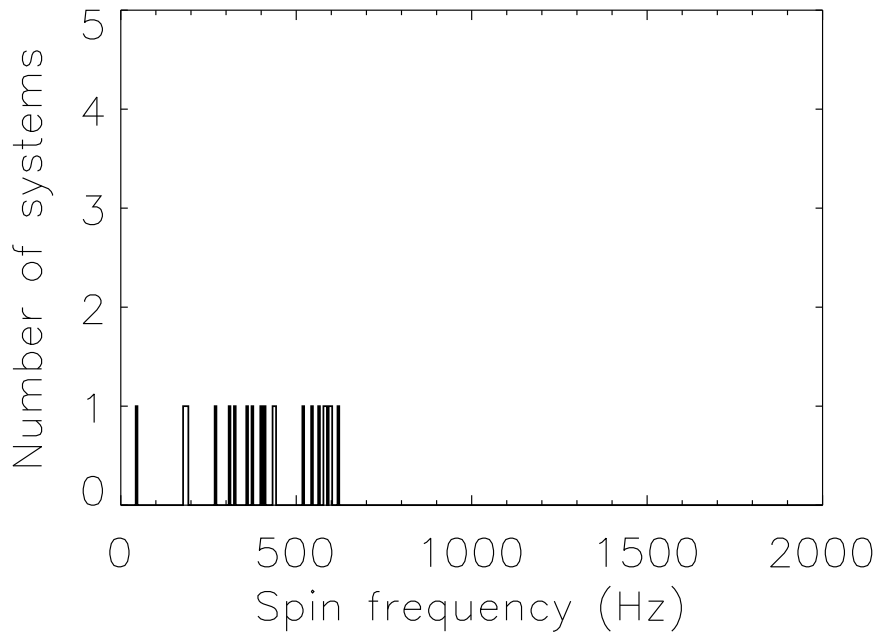}
  \caption{The spin frequency distribution of accreting millisecond
  X-ray pulsars.  There is a sharp cutoff in the population for spins
  above 730~Hz.  RXTE has no significant selection biases against
  detecting oscillations as fast as 2000~Hz, making the absence of
  fast rotators extremely statistically significant \citep{cmm+03,cha05}.}
\end{figure}

{\bf Measuring the spin distribution.}
The spin frequencies of the known AMXPs (both accretion-powered and
nuclear-powered) are plotted in Figure~1.  The spins are consistent
with a uniform distribution with the observed frequency range (45--619~Hz).
The absence of any pulsars above 619~Hz is extremely statistically
significant, given that there is no sigificant loss of RXTE
sensitivity out to at least 2~kHz.  Using only the subset of
nuclear-powered pulsars in order to minimize selection bias, and assuming
a uniform distribution up to some maximum frequency, the observed
distribution implies a maximum spin frequency of 730~Hz (95\%
confidence; \citep{cmm+03,cha05}). While the exact value for the
maximum spin frequency depends upon the particular prior assumption
chosen for the underlying distrbution, the existence of a 
cutoff frequency is robust.  However, the detailed shape of the
high-frequency distribution and the shape of the cutoff (flat out to
a sharp cutoff, ``pile up'' up to a sharp cutoff, tail of a Gaussian
distribution, etc.) cannot be constrained from the available sample,
although numerical simulations suggest that a mere factor of two
increase in sample size would already be revealing in this regard.

Assuming that the known AMXPs (or at least the nuclear-powered ones)
are a valid proxy for the general population of recycled millisecond
pulsars, it appears that recycled pulsars have a spin frequency limit
well below the centrifugal break-up rate for most equation of state
models.  Moreover, submillisecond pulsars are evidently very rare, if
they exist\footnote{There was a recent report of evidence for a
1122~Hz burst oscillation in XTE J1739$-$285 \citep{kpi+07}, but
independent analysis by several other groups (including ours)
indicates that the statistical significance of the reported signal is
marginal.}. The 716~Hz spin frequency of the fastest known millisecond
{\em radio} pulsar, PSR~J1748$-$2446ad in Ter~5 \citep{hrs+06}, is
consistent with these conclusions, as are the results of recent
attempts to infer the underlying spin distribution from radio pulsar 
observations \citep{mlc+05,cr05,hrs+07}. 

\section{HOW TO EXPLAIN THE SPIN DISTRIBUTION}

{\bf A very low breakup spin rate for neutron stars?}
It is unlikely that centrifugal breakup is the physics behind the
observed spin frequency cutoff. Figure~2 shows that a breakup
limit as low as 730~Hz excludes most equation of state models for
neutron stars, allowing only those with very large
radii \citep{cst94,lp01}. In order to accomodate the generally
accepted an 8--12~km radius range for a 1.4~$M_\odot$ neutron star
a breakup limit above $\sim$1500~Hz is required.   

\begin{figure}
  \includegraphics[height=.45\textheight]{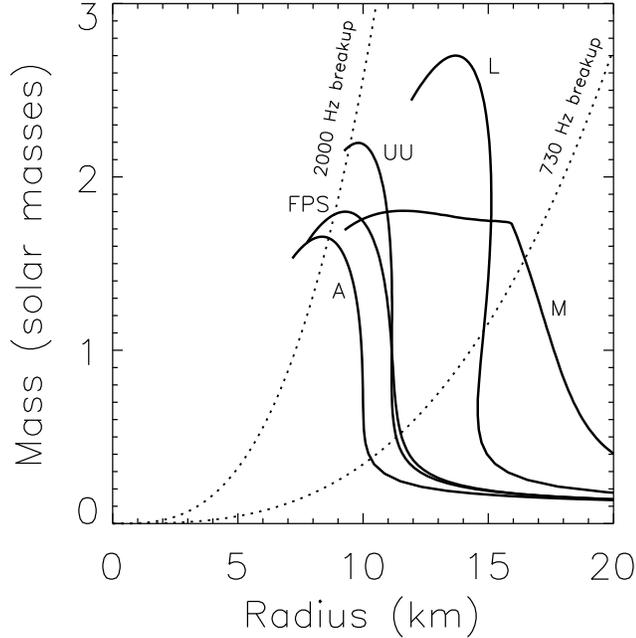}
  \caption{Neutron star break-up and equation of state.  The solid curves
  are theoretical mass-radius relations for a variety of models for
  the equation of state for ultradense matter from \citep{cst94}. The
  dashed curves show the limits arising from breakup spin rates of
  730~Hz and 2000~Hz; the allowed phase space is to the {\em right} of
  the appropriate dashed curve. We see that while a 2000~Hz breakup rate is
  consistent with most equations of state, a 730~Hz breakup rate
  is inconsistent with most models and excludes the 8--12~km radius
  range usually inferred for a 1.4~$M_\odot$ neutron star.}
\end{figure}

{\bf Pulsars not at magnetic spin equilibrium?}  One possible
alternative is that most of the systems in the sample are not at their
magnetic spin equilibrium.  While spin equilibrium is often assumed
for these systems since their accretion spin-up time scale is much
shorter than their X-ray active evolutionary lifetime, some authors
argue that many systems may not have reached equilibrium, or may
depart significantly from it on short time scales \citep{lb07}.  This
is primarily an evolutionary question that should properly be examined
on a case by case basis, involving the long-term accretion and
evolutionary history of the binary as well as issues of persistent
versus transient activity.  One could attempt to evaluate these issues
using short-term accretion torque measurements as well as binary
evolution modeling of individual systems.  However, even allowing for
departures from spin equilibrium on time scales much shorter than the
spin-up time, the basic time scale argument given above suggests that,
for a sufficiently large sample, a statistical assumption of spin
equilibrium should be valid.  Since many of the nuclear-powered AMXPs
are persistent X-ray sources, it is likely that at least significant
number of these systems is at or near spin equilibrium.

{\bf Pulsars are at magnetic spin equilibrium?}  Another alternative
is that the observed spin distribution simply reflects the
distribution of magnetic field strengths and mass accretion rates for
the observed systems, as related through the spin equilibrium
equation.  This is certainly the most conservative assumption (in that
it invokes no new physics beyond what is already used to explain X-ray
pulsations), but it requires that the AMXPs all have magnetic field
strengths of order $\sim 10^8$~G.  However, fields this strong should
be dynamically important and lead to magnetically channeled pulsations
from the persistent emission from all the systems.  Nonetheless, most
of the burst oscillation sources do {\em not} show detectable
pulsations in their non-burst emission.  While this might be due to
the same sort of intermittency effects observed in
Aql~X-1 \citep{cap+08}, possibly arising from fluctuations in the
accretion flow geometry \citep{lbw+08}, this may also reflect a wider
range of effective magnetic field strengths, with many systems not
having dynamically important fields \citep{czb01,cmm+03,cha05}.  The
most direct way to address this question would be to develop some
means for accurately estimating magnetic field strengths in the burst
oscillation sources, possibly using burst oscillation drift time
scales \cite{cmm+03}.  However, a more detailed measurement of the
spin distribution (i.e. with a larger sample) may also help, since the
shape expected for spin equilibria spanning a narrow range of magnetic
field strengths should look different than one affected by some
competing torque mechanism besides accretion torques.

{\bf Gravitational wave emission?}
An intriguing last alternative is that accretion spin-up torques in
the AMXPs may compete with angular momentum losses due to
gravitational radiation \citep{wag84,bil98}.  A variety of mechanisms
for the emission of gravitational waves from rapidly rotating
accreting neutron stars have been discussed in the literature,
including $r$-modes \citep{wag84,aks99}, accretion-induced crustal
quadrupoles \citep{bil98,ucb00}, toroidal magnetic
fields \citep{cut02}, and magnetically confined ``
mountains'' \citep{mp05}. Irrespective of the specific emission mechanism,
an attractive aspect of the gravitational wave model is that the
resulting spin-down torque scales very steeply with spin frequency,
$\propto\Omega^5$.  Thus, this model provides a natural way to produce a
sharp cutoff in population at high spin frequency, while not affecting
the usual magnetic spin equilibrium picture at lower frequencies.  

\section{TESTING GRAVITATIONAL RADIATION TORQUES}

Another attractive aspect of the gravitational radiation model is the
possibility of testing this hypothesis through the direct detection of
gravitational waves.  Although the intrinsic gravitational wave strain
amplitude expected from any these pulsars is small
($h<10^{-26}$; \citep{bil98}), these sources would be continuous
sources of gravitational wave emission in a frequency band accessible
by current and planned ground-based gravitational wave interferometer
experiments like LIGO and VIRGO, with the expected frequency being an
order unity multiple of the spin frequency (with the exact factor
depending upon the specific mechanism).  Thus, one could, in
principle, integrate the gravitational wave signal to obtain a
detection, in stark contrast to the short-lived transient ``chirp''
signals expected from binary merger events.

However, the known X-ray timing properties of the AMXPs will limit the
ability to integrate a gravitational wave signal coherently.  For
example, let us consider the effect of an accretion torque near
magnetic spin equilibrium,
\[
\dot\nu = 4\times 10^{-14} \left(\frac{\dot M}{0.01 \dot M_{\rm Edd}}\right)
          \left(\frac{\nu}{600 \mbox{\rm Hz}}\right)^{-1/3}
          \mbox{\rm\ Hz s$^{-1}$},
\]
where $\dot M_{\rm Edd}$ is the Eddington mass accretion rate, and $\nu$
and $\dot\nu$ are the spin 
frequency and its derivative.  Assuming steady accretion, this
corresponds to a decoherence time scale of
\[
  \tau_{\rm dec} = \sqrt{\frac{1}{\dot\nu}} \approx 
         60 \left(\frac{\dot M}{\dot M_{\rm Edd}}\right)^{-1/2}
            \left(\frac{\nu}{600 \mbox{\rm
         Hz}}\right)^{-1/6} \mbox{\rm d} .
\]
For Sco X-1 (the brightest persistent low-mass X-ray binary and the
most promising gravitational wave target), the mean accretion rate is
$\sim 0.5 \dot M_{\rm Edd}$, corresponding to a decoherence time of
only $\tau_{\rm dec}\approx 10$~d, complicating long integrations.
Moreover, long-term monitoring of SAX J1808.4$-$3658 has shown rapid
orbital evolution \citep{hpc+08,dbr+08}; if this is typical of AMXPs,
then it will lead to significant uncertainties in the Doppler
corrections necessary for coherent integration of a gravitational wave
signal, unless there is contemporaneous X-ray timing of the target.  A
recent detailed assessment by Watts et al. \cite{wkb+08} concludes
that direct LIGO detection of the predicted gravitational wave signals
from AMXPs will be prohibitively difficult given the current state of
X-ray timing models for the source population and the current
sensitivity of the LIGO detectors.

However, indirect detection of gravitational wave emission may be
possible through X-ray timing!   Accretion-powered AMXPs with
pulsations detected at multiple transient outburst epochs may be used
to study the long-term spin and orbital evolution of the pulsar.
Since one is measuring the mean spin frequency at widely separated
epochs, this analysis is relatively unaffected by the pulse shape
variability that can interfere with short-term timing studies.  The
best case for this sort of study is SAX J1808.4$-$3658, which has been
observed in outburst by RXTE in 1998, 2000, 2002, and 2005.  Precise
X-ray timing of these outbursts has revealed that the pulsar is
steadily spinning down, and that most of the torque is applied during
X-ray quiescence when accretion is shut off \citep{hpc+08}.  Accretion
torques are clearly not responsible.  In this particular case, the
observed spin down is consistent with magnetic dipole torques due
simply to the magnetized neutron star's rotation, given the known
constraints on the magnetic field strength.  These robust 
constraints come from applying magnetic accretion torque theory to the
observed range of X-ray luminosities over which accretion-powered
pulsations are detected \citep{pc99,hpc+08}.  Thus, for this
401~Hz pulsar, gravitational wave torques are not required to explain
the observed long-term spin down.

However, if gravitational wave torques are responsible for limiting
accretion spin-up to $\sim 730$~Hz, then we would not expect any
significant gravitational wave torques at 401~Hz, given their very steep
($\sim\Omega^5$) spin frequency dependence; we note that (401 Hz/730
Hz)$^5\approx 0.05$.  On the other hand, one would expect
gravitational wave emission to become more important for significantly
faster rotators, in which case magnetic dipole torques alone would
presumably be inadequate to explain any observed long-term spin down.  

The ideal target is thus the 599~Hz pulsar IGR~J00291+5934, the
fastest known accretion-powered millisecond pulsar \citep{gmm+05}; we
note that (599 Hz/730 Hz)$^5\approx 0.37$, a factor of 7 improvement
over SAX J1808.4$-$3658.  This target has only been timed with RXTE
during a single outburst in 2004.  However, an analysis of archival
RXTE/ASM data suggests that the source has a recurrence time of $\sim
3.2$~yr \cite{rem04}, so that the source should go into outburst in
the very near future.  If this outburst can be observed with RXTE,
then precise comparison of the spin frequency between 2004 and the new
outburst will yield a crude estimate of (or limit on) the long-term
spin-down rate.  The magnetic field strength of the pulsar can be
tightly constrained by examining the dynamic range of luminosity over
which X-ray pulsations are detected \citep{pc99}, and this can be used
to compute the expected contribution of magnetic dipole torques to the
spindown rate just as in SAX J1808.4$-$3658 \citep{hpc+08}.  If the
observed spindown rate significantly exceeds the expected magnetic
dipole spindown value, then that would be strong indirect evidence
that gravitational wave emission is affecting the pulsar's spin
evolution.  This is probably the most promising avenue for testing
gravitational wave spin down in AMXPs for the forseeable
future\footnote{This talk was delivered in Amsterdam in April 2008.  A
new outburst of IGR J00291+5934 was detected in August
2008 \citep{csm+08}, but it had an unusually brief duration of only a
few days.  This may not be long enough to allow a very precise spin
frequency measurement for comparison with the 2004 value. However, the
data are being reduced as of this writing, and we hope to apply the
analysis discussed above in order to test the gravitational wave
model.}.


\begin{theacknowledgments}
It is a pleasure to thank Lars Bildsten, Luciano Burderi, Tiziana Di
Salvo, Duncan Galloway, Jake Hartman, Scott Hughes, Miriam Krauss,
Fred Lamb, Jinrong Lin, Craig Markwardt, Ed Morgan, Michael Muno,
Feryal \"Ozel, Juri Poutanen, Dimitrios Psaltis, Paul Ray, Tod
Strohmayer, Jean Swank, Michiel van der Klis, Anna Watts, and Rudy
Wijnands for useful discussions and collaborations.  I also thank Rudy
Wijnands and the other meeting organizers for inviting me to
Amsterdam.
\end{theacknowledgments}


\begin{thebibliography}{9}

\bibitem{wv98}
Wijnands, R. \& van der Klis, M. 1998, \emph{Nature}, \textbf{398}, 344

\bibitem{cm98}
Chakrabarty, D. \& Morgan, E.~H. 1998, \emph{Nature}, \textbf{398}, 346

\bibitem{szs+96}
Strohmayer, T.~E., Zhang, W., Swank, J.~H., Smale, A., Titarchuk, L.,
Day, C., \& Lee, U. 1996, \emph{ApJ}, \textbf{469}, L9

\bibitem{vsz+96}
van der Klis, M., Swank, J.~H., Zhang, W., Jahoda, K., Morgan, E.~H.,
Lewin, W.~H.G., Vaughan, B., \& van Paradijs,
J. 1996, \emph{ApJ}, \textbf{469}, L1

\bibitem{cmm+03}
Chakrabarty, D., Morgan, E.~H., Muno, M.~P., Galloway, D.~K.,
Wijnands, R., van der Klis, M., \& Markwardt,
C.~B. 2003, \emph{Nature}, \textbf{424}, 42

\bibitem{brs93}
Bradt, H.~V., Rothschild, R.~E., \& Swank, J.~H. 1993, \emph{A\&AS},
\textbf{97}, 355 

\bibitem{acr+82}
Alpar, M.~A., Cheng, A.~F., Ruderman, M.~A., \& Shaham, J. 1982, 
\emph{Nature}, \textbf{300}, 728

\bibitem{rs82}
Radhakrishnan, V. \& Srinivasan, G. 1982, \emph{Curr.Sci.}, \textbf{51}, 1096

\bibitem{alp09}
Alpar, M.~A. 2009, these proceedings (arXiv:0808.3485)

\bibitem{gcm+02}
Galloway, D.~K., Chakrabarty, D., Morgan, E.~H., \& Remillard,
R.~A. 2002, \emph{ApJ}, \textbf{576}, L137

\bibitem{bdm+06}
Burderi, L., Di Salvo, T., Menna, M.~T., Riggio, A., \& Papitto, A. 2006,
\emph{ApJ}, \textbf{653}, 133

\bibitem{bdl+07}
Burderi, L. et al. 2007, \emph{ApJ}, \textbf{657}, 961

\bibitem{pmb+08}
Papitto, A., Menna, M.~T., Burderi, L., Di Salvo, T., \& Riggio, A. et
al. 2008, \emph{MNRAS}, \textbf{383}, 411 

\bibitem{rdb+08}
Riggio, A., Di Salvo, T., Burderi, L., Menna, M.~T., Papitto, A.,
Iaria, R., \& Lavagetto, G. 2008, \emph{ApJ}, \textbf{678}, 1273

\bibitem{gl79}
Ghosh, P. \& Lamb, F.~K. 1979, \emph{ApJ}, \textbf{234}, 296

\bibitem{pc99}
Psaltis, D. \& Chakrabarty, D. 1999, \emph{ApJ}, \textbf{521}, 332

\bibitem{hpc+08}
Hartman, J.~M., Patruno, A., Chakrabarty, D., Kaplan, D.~L.,
Markwardt, C.~B., Morgan, E.~H., Ray, P.~S., van der Klis, M., \&
Wijnands, R. 2008, \emph{ApJ}, \textbf{675}, 1468 

\bibitem{cst94}
Cook, G.~B., Shapiro, S.~L., \& Teukolsky,
S.~A. 1994, \emph{ApJ}, \textbf{421}, L117

\bibitem{hlz99}
Haensel, P, Lasota, J.~P, \& Zdunik,
J.~L. 1999, \emph{A\&A}, \textbf{344}, 
151

\bibitem{lp01}
Lattimer, J.~M. \& Prakash, M. 2001, \emph{ApJ}, \textbf{550}, 426

\bibitem{hrs+07}
Hessels, J.~W.~T., Ransom, S.~M., Stairs, I.~H., Kaspi, V.~M., \&
Freire, P.~C.~C. 2007, \emph{ApJ}, \textbf{670}, 363

\bibitem{gmk+07}
Galloway, D.~K., Morgan, E.~H., Krauss, M.~I., Kaaret, P., \&
Chakrabarty, D. 2007, \emph{ApJ}, \textbf{654}, L73

\bibitem{gss+07}
Gavriil, F., Strohmayer, T.~E., Swank, J.~H., \& Markwardt, C.~B. 2007,
\emph{ApJ}, \textbf{669}, L29

\bibitem{acp+08}
Altamirano, D., Casella, P., Patruno, A., Wijnands, R., \& van der
Klis, M. 2008, \emph{ApJ}, \textbf{674}, 45

\bibitem{cap+08}
Casella, P., Altamirano, D., Patruno, A., Wijnands, R., \& van der
Klis, M. 2008, \emph{ApJ}, \textbf{674}, 41

\bibitem{kpi+07}
Kaaret, P. et al. 2007, \emph{ApJ}, \textbf{657}, 97

\bibitem{hrs+06}
Hessels, J.~W.~T., Ransom, S.~M., Stairs, I.~H., Freire, P.~C.~C.,
Kaspi, V.~M., \& Camilo, F. 2006, \emph{Science}, \textbf{311}, 1901

\bibitem{mlc+05}
McLaughlin, M.~A. et al. 2005, in \emph{Binary Radio Pulsars},
ed. F.~A. Rasio \& I.~H. Stairs (ASP Conf. Proc. 328), 43

\bibitem{cr05}
Camilo, F. \& Rasio, F.~A. 2005, in \emph{Binary Radio Pulsars},
ed. F.~A. Rasio \& I.~H. Stairs (ASP Conf. Proc. 328), 147

\bibitem{cha05}
Chakrabarty, D. 2005, in \emph{Binary Radio Pulsars},
ed. F.~A. Rasio \& I.~H. Stairs (ASP Conf. Proc. 328), 279

\bibitem{lb07}
Lamb, F.~K., \& Boutloukos, S. 2007, in \emph{Short-Period Binary
Stars: Observation, Analysis, and Results}, ed. E.~F. Milone et al. 
(Springer), in press (arXiv:0705.0155)

\bibitem{lbw+08}
Lamb, F.~K. et al. 2008, \emph{ApJ}, submitted (arXiv:0808.4159)

\bibitem{czb01}
Cumming, A., Zweibel, E., \& Bildstenm L. 2001, \emph{ApJ}, \textbf{557}, 958

\bibitem{wag84}
Wagoner, R.~V. 1984, \emph{ApJ}, \textbf{278}, 345

\bibitem{bil98}
Bildsten, L. 1998, \emph{ApJ}, \textbf{501}, L89

\bibitem{aks99}
Andersson, N., Kokkotas, K.~D., \& Stergioulas,
N. 1999, \emph{ApJ}, \textbf{516}, 307

\bibitem{ucb00}
Ushomirsky, G., Cutler, C., \& Bildsten,
L. 2000, \emph{MNRAS}, \textbf{319}, 902 

\bibitem{cut02}
Cutler, C. 2002, \emph{Phys.Rev.D}, \textbf{66}, 4025

\bibitem{mp05}
Melatos, A. \& Payne, D.~J.~B. 2005, \emph{ApJ}, \textbf{623}, 1044

\bibitem{dbr+08}
Di Salvo, T., Burderi, L., Riggio, A., Papitto, A., \& Menna, M.~T. 2008,
\emph{MNRAS}, in press

\bibitem{wkb+08}
Watts, A.~L., Krishnan, B., Bildsten, L., \& Schutz, B.~F. 2008,
\emph{MNRAS}, \textbf{389}, 839

\bibitem{gmm+05}
Galloway, D.~K., Markwardt, C.~B., Morgan, E.~H., Chakrabarty, D., \&
Strohmayer, T.~E. 2005, \emph{ApJ}, \textbf{622}, L45

\bibitem{rem04}
Remillard, R. 2004, \emph{ATEL}, \textbf{357}

\bibitem{csm+08}
Chakrabarty, D., Swank, J.~H., Markwardt, C.~B., \& Smith, E. 2008,
\emph{ATEL}, \textbf{1660}


\end{thebibliography}
\end{document}